# Multi-wavelength Polarimetry of Isolated Neutron Stars

**Roberto P. Mignani**[1,2]

[1] INAF – Istituto di Astrofisica Spaziale e Fisica Cosmica, Milano, Via E. Bassini 15, 20133, Milan, Italy; mignani@iasf-milano.inaf.it
[2] Janusz Gil Institute of Astronomy, Unversity of Zielona Gora, ul. Szafrana 2, Zielona Gora, Poland
Academic Editor: Dr. Frédéric Marin

**Abstract:** Isolated Neutron Stars are known to be endowed with extreme magnetic fields, whose maximum intensity ranges from $10^{12}$ to $10^{15}$ G, which permeates their magnetospheres. Their surrounding environment is also strongly magnetised, especially in the compact nebulae powered by the relativistic wind from young neutron stars. The radiation from isolated neutron stars and their surrounding nebulae is, thus, supposed to bring a strong polarisation signature. Measuring the neutron star polarisation brings important information on the properties of their magnetosphere and of their highly magnetised environment. Being the most numerous class of isolated neutron stars, polarisation measurements have been traditionally carried out for radio pulsars, hence in the radio band. In this review, I summarise multi-wavelength linear polarisation measurements obtained at wavelengths other than radio both for pulsars and other types of isolated neutron stars and outline future perspectives with the upcoming observing facilities.

**Keywords:** X-rays; polarimetry; general; history of astronomy

## 1. Introduction

Over 2300 Isolated Neutron Stars (INSs), i.e. not in binary systems, have been identified so far according to the ATNF pulsar catalogue [1]. They represent a very important sample where we can study the multi-wavelength emission properties of the neutron star without being affected by the contribution of the companion star, which is dominating at wavelengths such as the optical, infrared, or ultraviolet. INSs are known in a variety of flavours, from the classical rotation-powered pulsars (RPPs), either radio-loud or radio-quiet, to the rotating radio transients (RRATs), to the magnetars, the most numerous INS class after the RPPs and powered by the magnetic energy, to the enigmatic Central Compact Objects (CCOs) in supernova remnants (SNRs), to the thermally emitting INSs (TINSs). The INS fraternity is discussed in the recent review of [2]. Their multi-wavelength phenomenology is very rich and diversified depending on the INS class, going from the RPPs, which can be detected across the entire electromagnetic spectrum, to the CCOs, which are only detected through their thermal emission in the soft X-rays. What makes one INS class different from another is one of the hotly debated issues in compact object astrophysics. Possible explanations point at a difference in the progenitor star, supernova explosion dynamics, or early evolution of the new-born neutron star. In some cases, differences in the intrinsic INS properties, such as the magnetic field strength, can help driving the investigations. A clear example is that of the CCOs and the magnetars, both in the same age range (a few kyrs), but with the former being characterised by a magnetic field strength a factor of $10^4$ lower than the latter.

It is obvious that multi-wavelength observations are key to characterise the INS diversity and determine their cause, possibly placing them in a unique evolutionary path. Polarimetry observations are one of the most important diagnostic tools since they provide, among other things, important information on the INS magnetosphere properties, allowing one to test different models of NS magnetosphere emission, including the location of the emission regions, and on the INS highly magnetised environment, including their pulsar wind nebulae (PWNe). Since radio pulsars are the most numerous class of INSs, polarisation measurements have been traditionally carried out in the radio band, also owing to the difficulties in carrying out polarisation measurements at other energies. These arise from the lower photon statistics with respect to the radio band (polarisation measurements require high counting rates), from the limited availability of cutting edge detectors, and, in the *gamma* and X-rays, from the lack of suitable instruments flying on the majority of space missions. As a matter of fact, apart from the radio band, optical polarimetry observations have been most successful in the study of INSs, exploiting a well-tested technology. In addition, optical observations represented the only way to measure the polarisation of INSs that are not detected in the radio band, such as the TINSs, or show, at most, sporadic radio emission, such as the magnetars. It is clear that the information obtained from multi-wavelength polarimetry, from the optical to the *gamma*-rays, complements that obtained from radio polarimetry alone, as it explores incoherent emission processes as opposed to the coherent ones. For this reason, multi-wavelength polarimetry of INSs is crucial for a global understanding and characterisation of these objects.

In this review, I will describe the state of the art of multi-wavelength polarimetry observations for different types of INSs carried out outside the radio band and I will outline future perspectives from the use of new observing facilities that are expected to become operational in the next decade.

## 2. Rotation-powered pulsars

As we said above, RPPs represent the most numerous class of INSs and, as such, they offer an ample choice of targets for polarimetry observations. Most importantly, RPPs are the only INS class for which several objects have been observed in different energy bands or even across the entire electromagnetic (non-radio) spectrum, from the infrared (IR), up to the very high-energy gamma-rays. For this reason, they are ideal targets for multi-wavelength polarisation studies. Indeed, RPPs are the only INS class for which polarisation measurements in at least one of these energy bands have been obtained for a few objects and the only INS class for which polarisation measurements in multiple energy bands have been obtained for at least one object (a notable case: the Crab pulsar).

### 2.1. RPP Optical polarimetry

The energy band where most non-radio polarisation measurements of RPPs have been obtained so far is the optical band (3000–8000 Å), an important channel to study radio-quiet RPPs. Being the first RPP detected in the optical [3] as well as the brightest (V=16.5), the Crab pulsar (PSR B0531+21) has been the first one for which a measurement of the optical linear polarisation[1] has been obtained [4] and the only one for which such measurements, both phase-averaged and phase-resolved, have been obtained through the years with a variety of telescopes/instruments (see, [5] and references therein). Phase resolved observations (e.g., [6]) have shown that the polarisation degree (PD) varies with the phases of the pulsar optical light curve and is maximum in the so-called bridge (BR) region (~25%), the phase interval between the primary and secondary pulse peaks (P1 and P2), and in the off-pulse (OP) region (~35%), the phase interval before the primary peak and after the secondary one, whereas it is minimum in coincidence with the two peaks (a few per cent). This suggested the presence of a strong DC component, possibly not associated with the pulsar. A similar behaviour is observed in the near-ultraviolet (UV) thanks to HST phase-resolved polarimetry observations [7], suggesting that the polarisation properties in the Crab are not wavelength-dependent, at least in the relatively narrow wavelength range between the optical and near-UV. Why the similarity in the polarisation properties between the optical and the near-UV does not reflect the difference between the optical/near-UV spectrum and light curves [8,9], as one would expect by assuming that the phase-resolved polarisation properties depend both on the pulsar spectrum and on the geometry of the emission region, is still to be understood. The optical phase-averaged PD= 9.8%±0.1% [6], which becomes 5.5%±0.1% after subtracting the DC component. Such a component was suggested [6] to be associated with the compact emission knot seen 0.6 arcsec from the pulsar, resolved by high-spatial resolution imaging with the Hubble Space Telescope (HST) but not in ground-based polarimetry observations. This interpretation has been confirmed by HST polarimetry observations [10,11] which measured a PD of 59.0%±1.9% for the knot against a value of 5.2%±0.3% measured for the pulsar, perfectly consistent with what measured from ground-based observations after accounting for the DC component. Quite interestingly, HST observations [11] have shown that the phase-averaged polarisation position angle (PA=105.1°±1.6°) was approximately aligned with the direction of the pulsar proper motion and with the axis of symmetry of the PWN observed in X-ray by Chandra. A similar alignment was also claimed for the Vela pulsar (PSR B0833-45), based upon ground-based observations [12], now confirmed by HST imaging polarimetry measurements [13], and interpreted as evidence of magneto-dynamic interaction between the pulsar and its surrounding environment. The Crab is also the only RPP for which the detection of optical circular polarisation has been announced [14], although not yet reported in a publication.

---

[1] In this manuscript polarisation has to be intended as linear polarisation unless stated otherwise.

As of now, there are five RPPs for which a measurement of the optical polarisation degree has been obtained, albeit with different level of significance owing to the different object brightness (see,[15] for a recent compilation and reference list). After the Crab, a robust measurement (8.1%±0.7%) has been obtained for the 11 kyr-old Vela pulsar (V=23.6), with the HST [13]. For the second brightest RPP identified in the optical, the 1.7 kyr-old PSR B0540-69 (V=22.5), the polarisation degree has also been obtained with the HST, albeit with discrepant values (5% and 16%), probably owing to a different approach in the subtraction of the PWN contamination and of the foreground polarisation along the Large Magellanic Cloud distance (49 kpc). For another young pulsar, PSR B1509-58 (1.7 kyr) only a tentative measurement, 10.4% with no reported error, has been published so far. PSR B0656+14 (V=25) is the only pulsar, apart from the Crab, for which both phase-averaged (11.9%±5.5%) and phase-resolved polarisation measurements have been obtained and is also the oldest and faintest one of the sample.

From a general point of view, one can see that the optical polarisation varies between 5% and 10% and is much lower than measured in the radio bands, probably owing to the different emission mechanisms. The observed optical polarisation is also a factor of few lower than expected from predictions based on most pulsar magnetosphere models, although such discrepancy can be due to over-simplification in the simulation codes or larger than expected depolarisation effects in the pulsar magnetosphere (see, [12] for a discussion). Bearing its current limitations in mind, the starting sample can be investigated for possible correlations between the polarisation degree and the pulsar properties [15]. For instance, a hint of a correlation between PD and the pulsar spin-down age, and of anti-correlation with the spin-down energy, can be seen, suggesting that the PD is higher for older and ess energetic pulsars. At the same time, a possible anti-correlation is seen between and the magnetic field at the pulsar light cylinder. Tentative as they might be, these hints can be biased by the fact that PSR B0656+14 is the only pulsar in the sample to occupy certain ranges of the parameter space. A more significant measurement of its PD, as well measuring the PD for other pulsars with similar characteristics (e.g., Geminga) is needed to assess the reality of the observed trends. Apparently, there is no dependence of the PD on the pulsar spectral index, suggesting that the way the optical radiation is generated does not obviously affect the level of polarisation it can achieve.

Although the current sample is a starting point for future observations, it has to be improved both in quality and quantity, passing through the revision of uncertain cases, such as PSR B1509 58. This is however, a challenging task owing to the object faintness (R~26) and its proximity (~0.6 arcsec) to a four magnitude brighter star, which requires both an 8 m-class telescope and a high spatial resolution in the optical. Furthermore, the pulsar is at low latitude with respect to the Galactic plane and at a distance of 4.4 kpc, so that the contribution of the foreground polarisation is expected to be dominant. Next best target for phase-averaged optical polarimetry is Geminga (V=25.5), now observed with the VLT (PI: Mignani), especially because of its close distance (<0.3 kpc), which minimises the effects of foreground polarisation, and it is not surrounded by an optical PWN, which minimises the polarisation background. Imaging polarimetry of Geminga would also be important to determine whether the alignment between the phase-averaged PD vector and the proper motion direction, seen in the Crab and Vela pulsars [11,13] and, possibly, in PSR B0656+14 [15], is an intrinsic characteristic of RPPs. Being much more informative than phase-average polarisation, phase-resolved polarisation should be pursued for the brightest targets, i.e. PSR B0540-69 and the Vela pulsar, using recently developed high time polarimetry instruments. Phase-resolved polarimetry would also be crucial to search for possible variations of the pulse PD occurring in coincidence with giant optical pulses, erratic (few per cent) variations of the single pulse intensity so far observed in the Crab pulsar only [16,17], and repeating almost exactly in phase with the giant radio pulses, hinting at a connection between incoherent and coherent emission. Both the origin and the connection of Giant optical and radio pulses in the Crab are still unclear but coordinated polarimetry observations in both bands can help to solve the mystery by tracing possible variations in the pulsar magnetosphere properties down to sub milli-second time scales. Since giant pulses have not yet been detected at X and gamma-ray energies (see, e.g. [18] and references therein), it is obvious that optical observations are pivotal to this goal.

*2.2. RPP X-ray polarimetry*

Polarisation observations of the Crab Nebula in the X-rays have been carried out since the early 1970s, with the first measurement yielding PD=15.4%±5.2% in the 5–20 keV energy range [19]. A similar but more precise value was obtained by [20], PD=15.7% ±1.5% with the OSO-8 satellite, which was revised to 19.2%±1.0% after subtraction of the pulsar contribution [21]. No new measurement has been tried/obtained in the next four decades, mainly owing to the fact that no X-ray polarimeter was flown on the on-orbit or planned X-ray satellites as a result of mission descoping or cancellation, budget cuts, or technical/scientific merit (see Weisskopf, these proceedings for an historical summary). Indeed, with only one object detected, X-ray polarimetry was considered a very risky enterprise, especially in view of the large investment in observing time.

Very recently, polarisation measurements of the Crab nebula in the 100–380 keV energy range have been obtained using the Cadmium Zinc Telluride Imager (CZTI) on board the Indian X-ray satellite ASTROSAT [22]. The phase-averaged PD was found to be 32.7%±5.8%, with a PA of 143.5°±2.8°. When phase-resolved, these measurements have shown that the PD is maximum in the OP region (up to ~80%) and is still quite high in the BR region (~20%). These results are pretty much in line with what has been already observed in the optical (see, e.g. [6] and references therein), where the PD is also maximum in the OP and BR regions. In the optical, it has been proposed [6,11] that the high PD measured in the phase interval away from the two peaks is associated with the knot seen close to the pulsar, which is very highly polarised (~59%). It is not clear whether the high PD observed in the X-rays in the same phase intervals is associated with an emission feature in the inner PWN, such as the knot itself which is unresolved at the Chandra spatial resolution [23], or it is intrinsic to the pulsar. A polarisation measurement of the Crab pulsar plus nebula system has been recently obtained with the Pogo+ ballon-borne X-ray polarimeter [24] in a close energy range (20–160 keV). Such measurement could only yield upper limits on the PD of the primary and secondary peaks (73% and 81%, respectively) and of the OP region (37%). The phase-averaged PD, however, was found to be 20.9%±5.0%, with a corresponding PA of 131.3°±6.8°, likely dominated by the nebula contribution. For no other RPP has the polarisation been measured in the X-rays.

*2.3. RPP Gamma-ray polarimetry*

The first polarisation measurements of the Crab in the soft *gamma*-ray regime (0.1–1 MeV) were obtained by the Integral satellite. Using the SPI detector [25] measured a PD of 46%±10% in the OP region with a PA=123°±11°, roughly aligned with the symmetry axis of the PWN. Phase-resolved *gamma*-ray polarisation measurements (0.2–0.8 MeV) were also obtained with IBIS, another instrument aboard Integral [26], yielded a PD of $42^{+30}_{16}$% for the two peaks and of > 88% for the OP plus BR regions. This showed that, like in the optical [6] and in the X-rays [22], the peaks seems to be much less polarised than the OP plus BR regions. A measurement of the *gamma*-ray polarisation in a similar energy range (0.13–0.44 MeV) was also obtained by [27] using Integral/SPI data.

Recently, [28] reported on a possible correlated variation of the Crab polarisation position angle both in the optical and in the soft *gamma*-rays by comparing multi-epoch measurements obtained with the HST and the Galway Astronomical Stoke Polarimeter (GASP; [29]) at the 5m Hale telescope and the Integral/IBIS instrument. More interestingly, these variations seemed to occur in coincidence with high-energy *gamma*-ray flaring events from the Crab Nebula detected by the Fermi satellite. The trend seems to be confirmed by new observations with IBIS and GASP at the 4.2 m William Herschel Telescope [30]. However, GASP observations, let alone the IBIS ones, had not the spatial angular resolution needed to resolve the pulsar from the nearby knot, s that it not possible to ascertain which of the two objects was responsible for the observed polarisation PA variability. Quite remarkably, no significant variation of the PD was noticed in parallel to those of the polarisation PA [28]. Whether such variations are indeed associated with *gamma-ray* flaring events in the Crab Nebula and whether they are associated either with the pulsar or the knot can be only ascertained by high spatial resolution monitoring polarimetry observations with the HST.

No measurement of the polarisation in the high-energy *gamma*-rays (> 50 MeV) has been obtained so far neither for the Crab nor for any other RPP. In a recent work, however, [31] studied the sensitivity to *gamma*-ray polarisation measurements using data from the Fermi Large Area Telescope (LAT) and found that with ten years of data the LAT could measure a minimum detectable polarisation (MDP) of 30%–50% at the $5\sigma$ level for the Crab and Vela pulsar. With Fermi now in its tenth year of operations, and continuing its mission till the end of the decade at least, this is certainly a possibility to be explored in the near future.

*2.3. RPP Multi-wavelength polarimetry*

The Crab is the only RPP for which polarisation measurements in two, or more, energy bands have been obtained. Some of the most representative values are summarised in Table 1. In most cases, the measurements have been phase-averaged so that it is difficult to separate the contribution of the pulsar from that of the background (the PWN). Only in the optical it has been possible to obtain a neat measurement of the pulsar PD without contamination from nearby sources (i.e. the knot) or the PWN, thanks to the exquisite spatial resolution of the HST. Comparing the PD obtained through different measurements is scientifically interesting but it is not straightforward owing to possible observational biases. First of all, owing to the different spatial resolution of the different instruments, such measurements encompass different areas around the pulsar, so that there are different polarisation background contributions from the PWN and the SNR.

Phase-resolved polarisation measurements can provide a direct measurement of the pulsar PD but they depend on how the BR and OP phase intervals are defined, which might be different from case to case, and rely on the assumption that the pulsar emission is totally pulsed. Finally, being the Crab PWNe an highly dynamic environment, the PD might be variable in time, and so its background contribution to the pulsar polarisation. Even with these caveats, one might speculate whether the PD is a function of energy. Indeed, looking at the values in Table 1 a hint of a trend for an increase of the PD with energy can be appreciated (see also Figure 4 of [24]), with the ASTROSAT measurement of [22] consistent with such a trend. New measurements, possibly unaffected by observational biases, would be fundamental to confirm such a trend, which, if real, would have important implications on the models of multi-wavelength emission from the pulsar magnetosphere. More difficult is to recognise a trend in PA, although it seems, on average, larger at higher energies than in the optical. In all cases, however, the PA values are more or less close to the axis of symmetry of the PWN but not perfectly aligned to each other. Whether the difference in the phase-averaged PAs may reflect a difference in the location of the emission regions and/or in the emission geometry in the NS magnetosphere is unclear.

| Energy range | Phase | PD (%) | PA (°) | Ref. |
|---|---|---|---|---|
| gamma-ray (0.1–1 MeV) | OP | 46 ±10 | 123 ±11 | [25] |
| gamma-ray (0.2–0.8 MeV) | P1+P2 | $42^{+30}_{-16}$ | 70±20 | [26] |
| gamma-ray (0.2–0.8 MeV) | OP | > 72 | 120.6± 8.5 | [26] |
| gamma-ray (0.2–0.8 MeV) | OP+BR | > 88 | 122.0± 7.7 | [26] |
| gamma-ray (0.2–0.8 MeV) | avg | $47^{+19}_{-13}$ | 100 ±11 | [26] |
| gamma-ray (0.13–0.44 MeV) | avg | 28 ±6 | 117 ±9 | [27] |
| X-ray (2.6 keV) | avg | 15.7±1.5 | 161.1±2.8 | [20] |
| X-ray (2.6 keV)[0] | avg | 19.2±1.0 | 156.4 ±1.4 | [21] |
| X-ray (20–120 keV) | avg | 20.9±5.0 | 131.3 ±6.8 | [24] |
| X-ray (100–380 keV) | avg | 32.7±5.8 | 143.5 ±2.8 | [22] |
| Optical[1] | avg | 9.8±0.1 | 109.5 ±0.1 | [6] |
| Optical[2] | avg | 5.5±0.1 | 96.4±0.1 | [10] |
| Optical[2] | avg | 5.2±0.3 | 105.1 ±1.6 | [11] |

[0] Nebula
[1] Pulsar plus knot
[2] Pulsar

## 2. Cooling INSs

While polarisation measurements of young RPPs, which are mostly sources of non-thermal radiation, allow one to obtain important information on the neutron star magnetosphere and on the surrounding PWNe, polarisation measurements of TINSs allow one to peak directly at, or close to, the NS surface, where their thermal radiation is produced as a result of the star cooling process. In particular, polarisation measurements can help addressing several open points on the emission mechanisms of these sources. First of all, it is not clear whether the thermal radiation comes from the bare NS surface or it is mediated by the presence of an atmosphere. It is also not clear what the composition of such atmosphere would be and whether and how much it would be magnetised. Furthermore, polarisation measurements are also an important tool to test the effects of quantum electrodynamics (QED), which are expected to manifest close to the NS surface. In particular, vacuum birefringence is expected to increase the level of polarisation from the NS surface from a few per cent up to 100 per cent, depending on the viewing geometry and the radiation emission mechanism ([32] and references therein). Thanks to their purely thermal radiation spectrum, TINSs are best suited to test QED effects. Thermal radiation from TINSs is detected in the soft X-rays and in the optical/UV, presumably from regions at different temperature on the NS surface. However, while their X-ray brightness, opposed to their optical/UV faintness, would indicate X-ray polarimetry as the most efficient way, no polarimeter working in the soft X-ray range is flying on any of the current X-ray satellites. Therefore, at variance with RPPs, polarimetry measurements of TINSs are only limited to the optical/UV.

Recently, [33] observed the TINS RX J1856.5-3754 using an optical imaging polarimeter at the VLT. RX J1856.5-3754 is the brightest (or less faint) INS of this class (V=25.5) and its optical spectral fluxes are best-fitted by a Rayleigh-Jeans, confirming the thermal nature of its radiation. Moreover, it is the closest one, with a parallactic distance of 123 pc only, which minimises the effect of foreground polarisation. The measured optical PD=16.43%±5.26%, which makes RX J1856.5-3754 the faintest INSs for which significant optical polarisation has been measured. Unfortunately, the still large statistical errors (much larger than the systematic ones) do not allow yet to discriminate between different emission models based upon the comparison between the predicted and simulated value [33]. Nonetheless, none of the tested thermal emission models (a black body, a magnetised, completely ionised hydrogen atmosphere, and a condensed surface) can account for the observed PD unless Nonetheless, none of the tested thermal emission models (a black body, a magnetised, completely ionised hydrogen atmosphere, and a condensed surface) can account for the observed PD unless QED vacuum birefringence effects are included in the simulations, which was assumed by [33] as an indirect evidence that vacuum birefringence indeed plays a role in determining the actual value of the PD (Figure 1a,b). Follow-up VLT observations (PI: Mignani) with the same instrument set-up have been collected in 2017 for a twice as long integration time and are being analysed while this manuscript is being written and should provide both an higher significance measurement of the PD and an independent confirmation of the vacuum birefringence effect. RX J1856.5 -3754 is, so far, the first and only TINS for which a measurement of the optical polarisation has been obtained.

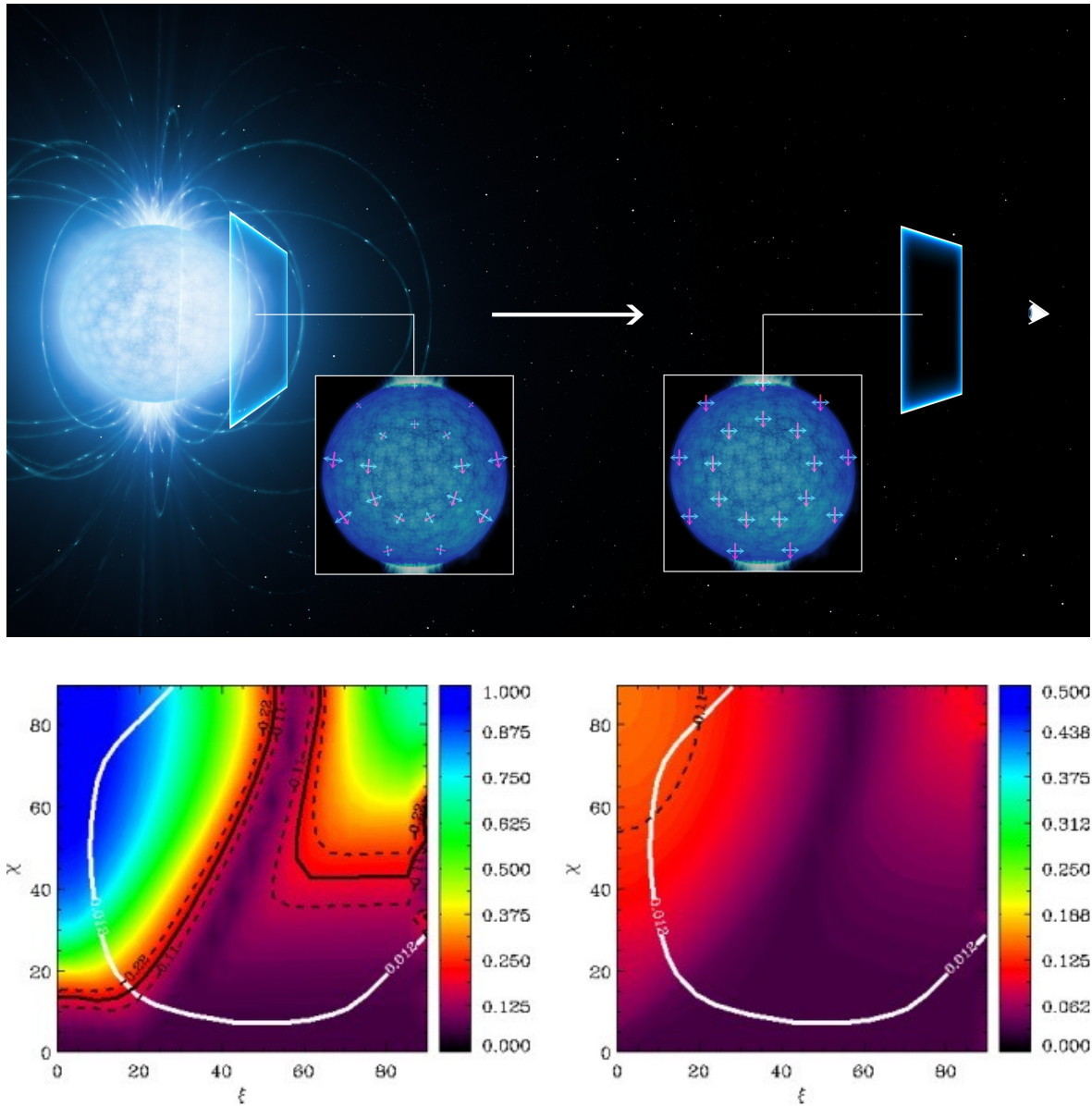

**Figure 1.** (**a**) Simulation of how the magnetic field affects the light polarisation as it propagates away from the neutron star, as predicted by the vacuum birefringence effect. In this case a distance up to 9 neutron star radii has been considered. The blue and red arrows indicate the electric and magnetic field vectors, respectively (image credit ESO). As it can be seen, the vector orientation changes with the distance from the neutron star surface, and so does the light polarisation. (**b**) Simulated PD of RX J1856.5-3754 (in colour scale) as a function of the angles $\chi$ and $\zeta$ i.e. between the line of sight and the INS spin axis and between the INS magnetic and spin axis, respectively. A black body emission model is assumed in the case shown here. The white solid line represents the constraints on $c$ and $x$ inferred from the pulsed X-ray light curve (see [33] for details), whereas the black solid line corresponds to the VLT optical polarisation measurement [33], including 1 $\sigma$ errors (black dashed line). The plots on the left and on the right represent the cases when QED effects are and are not accounted for, respectively (adapted from [33]).

## 2. Magnetars

Magnetars are the only remaining class of INSs for which polarisation measurements outside the radio bands have been obtained in at least one case. Being extremely magnetised objects, with magnetic fields up to three orders of magnitudes above that of the Crab pulsar, the radiation from the magnetar magnetosphere is expected to be highly polarised. Furthermore, magnetars are the most variable class of INSs, featuring bursts, outbursts, and giant flares, when the flux can vary up to several orders of magnitude (see, [34] for a review). Therefore, they are spectacular laboratories where to test the magnetic field evolution after or during one of such events and polarimetry represents one of the most useful diagnostic tools. Having a thermal X-ray spectrum harder than the TINSs, with the addition of non-thermal X-ray components detected up to 100 keV, magnetars can be targets of X-ray polarimetry observations. However, no measurement has been obtained with any of the current hard X-ray observing facilities, such as Integral.

Magnetars, however, have also been detected in the optical/IR, where the emission is non-thermal and, possibly, powered by the magnetic field, as suggested by their higher fraction of their spin-down power converted into optical/IR luminosity with respect to RPPs [35]. While they are usually intrinsically very faint in quiescence, and affected by a large interstellar extinction owing to their location on the Galactic plane and distances of a few kpc, magnetars can become relatively bright when they are in outburst. Phase-averaged imaging polarimetry observations in the IR have been carried out for some magnetars using the VLT (Israel et al. in prep.). For both 1E 1048-5937 and XTE J1810-197 only an upper limit of 25% on the PD has been obtained. For a third magnetar, 1E 1547.0-5408, follow-up observations of its IR counterpart, identified after an outburst onset [36], yielded a PD value of 4.2%±0.9%. This is the first measurement of a magnetar PD outside the radio band. However, the contribution of the foreground polarisation at the magnetar 5 kpc distance ($A_K$~1.9) is the largest source of uncertainty, which makes this value still preliminary. If correct, the measured PD might sound somehow lower than one would expect for an highly-magnetised object, like 1E 1547.0-5408 (B = 2.2 $10^{14}$ G). However, at least in the case of RPPs, no dependence has been found between the PD and the NS surface magnetic field [15], although the explored magnetic field range is relatively narrow (B = 3.4 $10^{12}$–1.54 $10^{13}$ G). The PD measured for 1E 1547.0-5408 is somewhat lower than the average for RPPs, which has been obtained in the optical (V band). No polarisation measurement in the IR has been obtained yet for any RPP which can be taken for comparison, though, so that we cannot rule out that the PD is wavelength-dependent and lower in the IR than in the optical. An upper limit on the circular polarisation of 4.3% has been obtained in the optical for the magnetar 4U 0142+61 with the Subaru telescope [37] and is the only one reported so far for an INS of this type.

## 3. Future Perspectives

In total, there are seven INSs (five RPPs, one TINS, one magnetar) for which a polarisation measurement has been obtained outside the radio band, i.e. in the optical/UV/IR, and one (the Crab pulsar) with a polarisation measurement also in the X-rays/soft gamma-rays. In the following of this section I describe the future perspectives for observations of these INS classes at optical, X and gamma-ray wavelengths with the upcoming facilities.

*5.1. Optical*

Being the INS class with more identifications in the optical domain (about 15, including candidates; Mignani et al. 2016), polarisation measurements of RPPs is one of the goals to be pursued with the current facilities. ESO observing programmes to obtain phase-resolved polarisation measurements of PSR B0540-69 and the Vela pulsar at the 3.6 m telescope with the GASP instrument (PI: Shearer) and phase-averaged measurements of Geminga at the VLT (PI: Mignani) have been approved and the observations are now being carried out or completed. Follow-up phase-resolved observations for PSR B0656+14 and, possibly, Geminga should also been pursued depending on the possibility of hosting guest instruments, such as GASP, at 8m-class telescopes, either at Northern or Southern hemisphere telescopes. For some of the remaining targets, polarimetry observations are complicated by the proximity to nearby stars (e.g., PSR B1509-58, PSR B1055-52, and PSR J1751-2054), whereas for the faintest ones, with fluxes fainter than magnitude 26, we would require very large integrations (few tens of hours) with 8m-class telescopes and very good and stable seeing conditions in dark time to maximise the signal–to–noise, which are difficult to match in ground-based observatories. The same is true for the TINSs, which are all fainter than RX J1856.4-3754 in the optical. The perspective is a bit more promising for magnetars, which can be relatively bright in outburst, but polarimetry observations must rely on quick response time to observe the source at the brightness peak and still have to cope with the problem of foreground polarisation. In the forthcoming future, observations with the 30-40 m class telescopes, such as the ESO ELT, might provide the required collecting power and spatial resolution. However, at least according to current plans, no polarimeter should be among the first generation instruments. Moreover, in the case of the ELT, multiple reflections in the mirror system would increase the instrument polarisation, making it difficult to calibrate it out. The plans for the VLT after 2025, however, are still unclear but it is desirable that the advent of the ELT will trigger a rethinking of the scientific priorities of the Paranal Observatory, offering the opportunity to mount (semi-)permanently optical polarisation instruments. The HST is an ideal telescope for high-spatial resolution imaging observations and it would play a crucial role to look for changes of polarisation of the Crab pulsar plus knot system in coincidence with gamma-ray flares [28,30] disentangling the contribution of the two components and helping to solve the long-standing mystery on the origin of such events.

*5.2. X-rays*

With the Crab pulsar being the only INSs for which X-ray polarisation measurements have been obtained, it is obvious that X-ray polarimetry is a resource largely unexplored. Perspectives will change in the coming decade with the deployment of new observing facilities dedicated to X-ray polarimetry observations, such as NASA's Imaging X-ray Polarimeter Explorer (IXPE; [38]) a SMEX-class mission due to fly in the early 2020s, ESA's X-ray Imaging Polarimeter Explorer (XIPE; [39]), a candidate medium-class mission with a launch window in the mid 2020s[2], and the enhanced X- ray Timing Polarimetry mission (eXTP; [40]), a Chinese project with European partner institutions already approved by the Chinese Academy of Science and due to fly around 2025. These three mission are based on the Gas Pixel Detector (GPD) technology and have been designed to work in roughly the same energy range (2–8 keV or 2–10 keV), with a similar field of view ( 12 12 arcmin), with a time resolution better than 100 ms (XIPE's would be better than 8 ms), although with different effective areas, with that of both XIPE and eXTP being about twice as high that of IXPE. This translates to a minimum detectable polarisation (MDP) between  1% and  2% for the same flux limit ($2 \times 10^{10}$ erg /cm$^2$ s$^{-1}$=10 mCrab) and integration time (300 ks). Owing to the energy range, only RPPs and magnetars will be suitable targets for these missions (see Taverna and Turolla, these proceedings, for a discussion of the magnetar case). The brightest target RPPs are also embedded in bright X-ray PWNe, which represent a major source of background polarisation that cannot be easily subtracted owing to the coarse angular resolution of the GPD detectors (< 30 arcsec), which, in most cases, is larger than the angular dimension of the PWN. Moreover, X-ray PWNe are also known to vary in flux (e.g. the Crab and Vela

---

[2] At the time of writing, the official selection of the ESA medium-class mission for cycle 4 (M4) has been postponed to February 2018

PWNe) and so might be the polarisation background, although flux changes do not necessarily come together with changes in the polarisation level, at least this is the case in the optical [11]. While for compact pulsar/PWNe systems targets can be selected by requiring than the PWN X-ray flux does not exceed a certain fraction of the pulsar flux, it has to be noted that a low X-ray flux does not necessarily imply a low polarisation level. Therefore, since most of the best-suited target RPPs, selected on the basis of their 2–10 keV flux and corresponding MDP, are also X-ray pulsars in this energy band, the best approach would be to exploit the GPD time resolution to isolate the pulsar contribution (pulsed) from that of the PWN (unpulsed) through phase-resolved polarimetry. While no dedicated mission has been approved yet, concept studies for X-ray polarimeters working in the soft X-ray band (0.1–2 keV) are carried out, with prototypes to be tested on suborbital flights (see Marshall et al., these proceedings). Extending the sensitivity range of X-ray polarimetry would be fundamental to study TINSs, for which polarisation measurements in the optical domain are challenging owing to the target faintness. Measurements of the X-ray PD would be crucial to complement, and test on a larger sample, the results on vacuum birefringence already obtained in the optical [33] and verify the predicted dependence of this effect on the photon energy.

*5.3. Gamma-rays*

While Integral will be still operational for the next few years and will likely be used to continue its monitoring of the Crab pulsar in the soft *gamma*-rays, and Fermi might allow to measure polarisation in the high-energy *gamma*-rays at least for the Crab and Vela pulsar [31], it is clear that a major step forward in INS studies must rely on a new observing facility with breakthrough capabilities for *gamma*-ray polarimetry. The e-ASTROGAM satellite [41], a candidate mission for *gamma*-ray astrophysics in the 0.15 Mev–3 GeV energy range, now being evaluated for the ESA M5 selection, satisfy these requirements. e-ASTROGAM would be able to carry out polarisation measurements through the pair creation and Compton scattering techniques, allowing one to achieve an MDP as low as 1% in the low-energy range (0.2–2 MeV) for a Crab-like source with a 1 Ms integration [42]. Other RPPs detected at low-energy *gamma*-rays by the COMPTEL instrument aboard the Compton Gamma-ray Observatory would also be prime targets for polarimetry observations with e-ASTROGAM [43]. The PWN polarisation contamination problem would likely be more severe than in the X-rays and cannot be directly subtracted through background sampling owing to the coarser spatial resolution in the *gamma*-rays. Therefore, target selection criteria might be different and, presumably, mostly based on photon selection from the *gamma*-ray light curve analysis. Like in the X-rays, the advent of this mission would open the so far barely explored field of INS *gamma*-ray polarimetry.

**3. Conclusions**

Polarisation studies of INSs are still in their infancy, with very few objects (mostly RPPs) with a polarisation measurement outside the radio band, typically in the optical. This is especially true at high energies where only for the Crab pulsar it has been possible to measure polarisation both in the hard X-rays and in the soft gamma-rays. Even these very few detections, however, have shown the potentials of the diagnostic power of polarimetry in the field of pulsar electrodynamics, e.g. by testing INS magnetosphere and emission models, and in quantum physics, e,g. by verifying QED predictions, such as vacuum birefringence, which built a new bridge between the astrophysics and fundamental physics communities. Times are, thus, mature to carry out polarisation studies of INSs on a larger scale, capitalising on the momentum built in the astronomical community. The blossoming of proposed/planned observing facilities dedicated to, or capable of, polarimetry observations witnesses this change. These new facilities will expand by a factor of ten to twenty the number of INSs with detected polarised high-energy emission, opening scientific horizons unforeseen till now. From a general stand point, multi-wavelength polarisation measurements will allow one to study the properties of INS magnetic fields and magnetospheres in different energy regimes, verify the dependence of the polarisation level on the photon energy and the INS spectrum for the first time, and disentangle different emission processes, for which different polarisation signatures are expected. With IXPE, eXTP, and XIPE, in the X-rays, eASTROGAM in the gamma-rays, if approved, and hopefully dedicated instruments at future optical facilities (ELTs), in the mid 2020s we will enter the new era of multi-wavelength polarimetry, adding a fourth technique, after imaging, timing, and spectroscopy, for the multi-wavelength study of INSs.

**Acknowledgments:** The author acknowledges financial support from an INAF Occhialini Fellowship.

**Conflicts of Interest:** The authors declare no conflict of interest.

**References**

1. Manchester, R.N.; Hobbs, G.B.; Teoh, A.; Hobbs, M. The Australia Telescope National Facility Pulsar Catalogue. Astronomical Journal **2005**, 129, 1993–2006, [astro-ph/0412641].

2. Harding, A.K. The neutron star zoo. Frontiers of Physics **2013**, 8, 679–692, [arXiv:astro ph.HE/1302.0869].

3. Cocke, W.J.; Disney, M.J.; Taylor, D.J. Discovery of Optical Signals from Pulsar NP 0532. Nature **1969**, 221, 525–527.

4. Wampler, E.J.; Scargle, J.D.; Miller, J.S. Optical Observations of the Crab Nebula Pulsar. Astrophysical Journal **1969**, 157, L1.

5. Mignani, R. The Crab Optical and UV Polarimetry. Polarimetry days in Rome: Crab status, theory and prospects, 2008, p. 9.

6. Słowikowska, A.; Kanbach, G.; Kramer, M.; Stefanescu, A. Optical polarization of the Crab pulsar: precision measurements and comparison to the radio emission. Monthly Notices of the Royal Astronomical Society **2009**,397, 103–123, [arXiv:astro-ph.SR/0901.4559].

7. Graham-Smith, F.; Dolan, J.F.; Boyd, P.T.; Biggs, J.D.; Lyne, A.G.; Percival, J.W. The ultraviolet polarization of the Crab pulsar. Monthly Notices of the Royal Astronomical Society **1996**, 282, 1354–1358.

8. Sollerman, J.; Lundqvist, P.; Lindler, D.; Chevalier, R.A.; Fransson, C.; Gull, T.R.; Pun, C.S.J.; Sonneborn, G. Observations of the Crab Nebula and Its Pulsar in the Far-Ultraviolet and in the Optical. The American Astronomical Society **2000**, 537, 861–874, [astro-ph/0002374].

9. Percival, J.W.; Biggs, J.D.; Dolan, J.F.; Robinson, E.L.; Taylor, M.J.; Bless, R.C.; Elliot, J.L.; Nelson, M.J.; Ramseyer, T.F.; van Citters, G.W.; Zhang, E. The Crab pulsar in the visible and ultraviolet with 20microsecond effective time resolution. The American Astronomical Society **1993**, 407, 276–283.

10. Slowikowska, A.; Mignani, R.; Kanbach, G.; Krzeszowski, K. Decomposition of the Optical Polarisation Components of the Crab Pulsar and its Nebula. Electromagnetic Radiation from Pulsars and Magnetars; Lewandowski, W.; Maron, O.; Kijak, J., Eds., 2012, Vol. 466, Astronomical Society of the Pacific ConferenceSeries, p. 37.

11.  Moran, P.; Shearer, A.; Mignani, R.P.; Słowikowska, A.; De Luca, A.; Gouiffès, C.; Laurent, P. Optical polarimetry of the inner Crab nebula and pulsar. Monthly Notices of the Royal Astronomical Society **2013**, 433, 2564–2575, [arXiv:astro-ph.HE/1305.6824].

12.  Mignani, R.P.; Bagnulo, S.; Dyks, J.; Lo Curto, G.; Słowikowska, A. The optical polarisation of the Vela pulsar revisited. Astronomy and Astrophysics **2007**, 467, 1157–1162, [astro-ph/0702307].

13.  Moran, P.; Mignani, R.P.; Shearer, A. HST optical polarimetry of the Vela pulsar and nebula. Monthly Notices of the Royal Astronomical Society **2014**, 445, 835–844, [arXiv:astro-ph.HE/1409.1743].

14.  Wiktorowicz, S.; Ramirez-Ruiz, E.; Illing, R.M.E.; Nofi, L. Discovery of Optical Circular Polarization of the Crab Pulsar. American Astronomical Society Meeting Abstracts, 2015, Vol. 225, American Astronomical Society Meeting Abstracts, p. 421.01.

15.  Mignani, R.P.; Moran, P.; Shearer, A.; Testa, V.; Słowikowska, A.; Rudak, B.; Krzeszowski, K.; Kanbach, G. VLT polarimetry observations of the middle-aged pulsar PSR B0656+14. Astronomy and Astrophysics **2015**, A105, [arXiv:astro-ph.HE/1510.01057].

16.  Shearer, A.; Stappers, B.; O'Connor, P.; Golden, A.; Strom, R.; Redfern, M.; Ryan, O. Enhanced Optical Emission During Crab Giant Radio Pulses. Science **2003**, 301, 493–495, [astro-ph/0308271].

17.  Strader, M.J.; Johnson, M.D.; Mazin, B.A.; Spiro Jaeger, G.V.; Gwinn, C.R.; Meeker, S.R.; Szypryt, P.; van Eyken, J.C.; Marsden, D.; O'Brien, K.; Walter, A.B.; Ulbricht, G.; Stoughton, C.; Bumble, B. Excess Optical Enhancement Observed with ARCONS for Early Crab Giant Pulses. Astrophysical Journal **2013**, 779, L12, [arXiv:astro-ph.HE/1309.3270].

18.  Hitomi Collaboration.; Aharonian, F.; Akamatsu, H.; Akimoto, F.; Allen, S.W.; Angelini, L.; Audard, M.; Awaki, H.; Axelsson, M.; Bamba, A.; Bautz, M.W.; Blandford, R.; Brenneman, L.W.; Brown, G.V.; Bulbul, E.; Cackett, E.M.; Chernyakova, M.; Chiao, M.P.; Coppi, P.S.; Costantini, E.; de Plaa, J.; de Vries, C.P.; den Herder, J.W.; Done, C.; Dotani, T.; Ebisawa, K.; Eckart, M.E.; Enoto, T.; Ezoe, Y.; Fabian, A.C.; Ferrigno, C.; Foster, A.R.; Fujimoto, R.; Fukazawa, Y.; Furuzawa, A.; Galeazzi, M.; Gallo, L.C.; Gandhi, P.; Giustini, M.; Goldwurm, A.; Gu, L.; Guainazzi, M.; Haba, Y.; Hagino, K.; Hamaguchi, K.; Harrus, I.M.; Hatsukade, I.; Hayashi, K.; Hayashi, T.; Hayashida, K.; Hiraga, J.S.; Hornschemeier, A.; Hoshino, A.; Hughes, J.P.; Ichinohe, Y.; Iizuka, R.; Inoue, H.; Inoue, Y.; Ishida, M.; Ishikawa, K.; Ishisaki, Y.; Iwai, M.; Kaastra, J.; Kallman, T.; Kamae, T.; Kataoka, J.; Katsuda, S.; Kawai, N.; Kelley, R.L.; Kilbourne, C.A.; Kitaguchi, T.; Kitamoto, S.; Kitayama, T.; Kohmura, T.; Kokubun, M.; Koyama, K.; Koyama, S.; Kretschmar, P.; Krimm, H.A.; Kubota, A.; Kunieda, H.; Laurent, P.; Lee, S.H.; Leutenegger, M.A.; Limousin, O.O.; Loewenstein, M.; Long, K.S.; Lumb, D.; Madejski, G.; Maeda, Y.; Maier, D.; Makishima, K.; Markevitch, M.; Matsumoto, H.; Matsushita, K.; McCammon, D.; McNamara, B.R.; Mehdipour, M.; Miller, E.D.; Miller, J.M.; Mineshige,S.; Mitsuda, K.; Mitsuishi, I.; Miyazawa, T.; Mizuno, T.; Mori, H.; Mori, K.; Mukai, K.; Murakami, H.; Mushotzky, R.F.; Nakagawa, T.; Nakajima, H.; Nakamori, T.; Nakashima, S.; Nakazawa, K.; Nobukawa, K.K.; Nobukawa, M.; Noda, H.; Odaka, H.; Ohashi, T.; Ohno, M.; Okajima, T.; Oshimizu, K.; Ota, N.; Ozaki, M.; Paerels, F.; Paltani, S.; Petre, R.; Pinto, C.; Porter, F.S.; Pottschmidt, K.; Reynolds, C.S.; Safi-Harb, S.; Saito, S.; Sakai, K.; Sasaki, T.; Sato, G.; Sato, K.; Sato, R.; Sawada, M.; Schartel, N.; Serlemtsos, P.J.; Seta, H.; Shidatsu, M.; Simionescu, A.; Smith, R.K.; Soong, Y.; ukasz Stawarz, Ł.; Sugawara, Y.; Sugita, S.; Szymkowiak, A.; Tajima,

H.; Takahashi, H.; Takahashi, T.; Takeda, S.; Takei, Y.; Tamagawa, T.; Tamura, T.; Tanaka, T.; Tanaka, Y.; Tanaka, Y.T.; Tashiro, M.S.; Tawara, Y.; Terada, Y.; Terashima, Y.; Tombesi, F.; Tomida, H.; Tsuboi, Y.; Tsujimoto, M.; Tsunemi, H.; Tsuru, T.G.; Uchida, H.; Uchiyama, H.; Uchiyama, Y.; Ueda, S.; Ueda, Y.; Uno, S.; Urry, C.M.; Ursino, E.; Watanabe, S.; Werner, N.; Wilkins, D.R.; Williams, B.J.; Yamada, S.; Yamaguchi, H.; Yamaoka, K.; Yamasaki, N.Y.; Yamauchi, M.; Yamauchi, S.; Yaqoob, T.; Yatsu, Y.; Yonetoku, D.; Zhuravleva, I.; Zoghbi, A.; Terasawa, T.; Sekido, M.; Takefuji, K.; Kawai, E.; Misawa, H.; Tsuchiya, F.; Yamazaki, R.; Kobayashi, E.; Kisaka, S.; Aoki, T. Hitomi X-ray studies of Giant Radio Pulses from the Crab pulsar. ArXiv e-prints **2017**, [arXiv:astro-ph.HE/1707.08801].

19. Novick, R.; Weisskopf, M.C.; Berthelsdorf, R.; Linke, R.; Wolff, R.S. Detection of X-Ray Polarization of the Crab Nebula. Astrophysical Journal **1972**, 174, L1.

20. Weisskopf, M.C.; Cohen, G.G.; Kestenbaum, H.L.; Long, K.S.; Novick, R.; Wolff, R.S. Measurement of the X-ray polarization of the Crab Nebula. Astrophysical Journal **1976**, 208, L125–L128.

21. Weisskopf, M.C.; Silver, E.H.; Kestenbaum, H.L.; Long, K.S.; Novick, R. A precision measurement of the X-ray polarization of the Crab Nebula without pulsar contamination. Astrophysical Journal **1978**, L117–L121.

22. Vadawale, S.V.; Chattopadhyay, T.; Mithun, N.P.S.; Rao, A.R.; Bhattacharya, D.; Vibhute, A.; Bhalerao, V.B.; Dewangan, G.C.; Misra, R.; Paul, B.; Basu, A.; Joshi, B.C.; Sreekumar, S.; Samuel, E.; Priya, P.; Vinod, P.; Seetha, S. Phase-resolved X-ray polarimetry of the Crab pulsar with the AstroSat CZT Imager. Nature Astronomy **2018**, 2, 50–55.

23. Rudy, A.; Horns, D.; DeLuca, A.; Kolodziejczak, J.; Tennant, A.; Yuan, Y.; Buehler, R.; Arons, J.; Blandford, R.; Caraveo, P.; Costa, E.; Funk, S.; Hays, E.; Lobanov, A.; Max, C.; Mayer, M.; Mignani, R.; O'Dell, S.L.; Romani, R.; Tavani, M.; Weisskopf, M.C. Characterization of the Inner Knot of the Crab: The Site of the Gamma-Ray Flares? The American Astronomical Society **2015**, 811, 24, [arXiv:astro-ph.HE/1504.04613].

24. Chauvin, M.; Florén, H.G.; Friis, M.; Jackson, M.; Kamae, T.; Kataoka, J.; Kawano, T.; Kiss, M.; Mikhalev, V.; Mizuno, T.; Ohashi, N.; Stana, T.; Tajima, H.; Takahashi, H.; Uchida, N.; Pearce, M. Shedding new light on the Crab with polarized X-rays. Scientific Reports **2017**, 7, 7816, [arXiv:astro-ph.HE/1706.09203].

25. Dean, A.J.; Clark, D.J.; Stephen, J.B.; McBride, V.A.; Bassani, L.; Bazzano, A.; Bird, A.J.; Hill, A.B.; Shaw, S.E.; Ubertini, P. Polarized Gamma-Ray Emission from the Crab. Science **2008**, 321, 1183.

26. Forot, M.; Laurent, P.; Grenier, I.A.; Gouiffès, C.; Lebrun, F. Polarization of the Crab Pulsar and Nebula as Observed by the INTEGRAL/IBIS Telescope. Astrophysical Journal **2008**, 688, L29, [0809.1292].

27. Chauvin, M.; Roques, J.P.; Clark, D.J.; Jourdain, E. Polarimetry in the Hard X-Ray Domain with INTEGRAL SPI. The American Astronomical Society **2013**, 769, 137, [arXiv:astro-ph.IM/1305.0802].

28. Moran, P.; Kyne, G.; Gouiffès, C.; Laurent, P.; Hallinan, G.; Redfern, R.M.; Shearer, A. A recent change in the optical and *gamma*-ray polarization of the Crab nebula and pulsar. Monthly Notices of the Royal Astronomical Society **2016**, 456, 2974–2981, [arXiv:astro-ph.HE/1511.07641].

29. Collins, P.; Kyne, G.; Lara, D.; Redfern, M.; Shearer, A.; Sheehan, B. The Galway astronomical Stokes polarimeter: an all-Stokes optical polarimeter with ultra-high time resolution. Experimental Astronomy **2013**, 36, 479–503, [arXiv:astro-ph.IM/1305.6825].

30. Gouiffes, C.; Laurent, P.; Shearer, A.; O'Connor, E.; Moran, P. New hard X-rays and optical polarimetric observations of the Crab nebula and pulsar. Proceedings of the 11th INTEGRAL Conference Gamma-Ray Astrophysics in Multi-Wavelength Perspective. 10-14 October 2016 Amsterdam, The Netherlands (INTEGRAL2016). Online at , 2016, p. 38.

31. Giomi, M.; Bühler, R.; Sgrò, C.; Longo, F.; Atwood, W.B. Estimate of the Fermi large area telescope sensitivity to gamma-ray polarization. 6th International Symposium on High Energy Gamma-Ray Astronomy, 2017, Vol. 1792, American Institute of Physics Conference Series, p. 070022, [arXiv:astro-ph.IM/1610.06729].

32. Heyl, J.S.; Shaviv, N.J.; Lloyd, D. The high-energy polarization-limiting radius of neutron star magnetospheres - I. Slowly rotating neutron stars. Monthly Notices of the Royal Astronomical Society **2003**, 342, 134–144, [astro-ph/0302118].

33. Mignani, R.P.; Testa, V.; González Caniulef, D.; Taverna, R.; Turolla, R.; Zane, S.; Wu, K. Evidence for vacuum birefringence from the first optical-polarimetry measurement of the isolated neutron star RX J1856.5-3754. Monthly Notices of the Royal Astronomical Society **2017**, 465, 492–500, [arXiv:astro-ph.HE/1610.08323].

34. Turolla, R.; Zane, S.; Watts, A.L. Magnetars: the physics behind observations. A review. Reports on Progress in Physics **2015**, 78, 116901, [arXiv:astro-ph.HE/1507.02924].

35. Mignani, R.P.; Perna, R.; Rea, N.; Israel, G.L.; Mereghetti, S.; Lo Curto, G. VLT/NACO observations of the high-magnetic field radio pulsar PSR J1119-6127. Astronomy and Astrophysics **2007**, 471, 265–270, [arXiv:astro-ph.HE/0706.2573].

36. Israel, G.L.; Rea, N.; Rol, E.; Mignani, R.; Testa, V.; Stella, L.; Esposito, P.; Mereghetti, S.; Tiengo, A.; Marconi, G.; Burgay, M.; Possenti, A.; Zane, S. ESO-VLT discovery of the variable nIR counterpart to the AXP 1E1547.0-5408. The Astronomer's Telegram **2009**, 1909.

37. Wang, Z.; Tanaka, Y.T.; Zhong, J. Subaru Constraint on Circular Polarization in I-Band Emission rom the Magnetar 4U 0142+61. Publications of the Astronomical Society of Japan **2012**, 64, [arXiv:astro-ph.HE/1111.5104].

38. Weisskopf, M.C.; Ramsey, B.; O'Dell, S.; Tennant, A.; Elsner, R.; Soffitta, P.; Bellazzini, R.; Costa, E.; Kolodziejczak, J.; Kaspi, V.; Muleri, F.; Marshall, H.; Matt, G.; Romani, R. The Imaging X-ray Polarimetry Explorer (IXPE). Space Telescopes and Instrumentation 2016: Ultraviolet to Gamma Ray, 2016, Vol. 9905, Proceedings of SPIE, p. 990517.

39. Soffitta, P.; Bellazzini, R.; Bozzo, E.; Burwitz, V.; Castro-Tirado, A.; Costa, E.; Courvoisier, T.; Feng, H.;bGburek, S.; Goosmann, R.; Karas, V.; Matt, G.; Muleri, F.; Nandra, K.; Pearce, M.; Poutanen, J.; Reglero, V.; Sabau Maria, D.; Santangelo, A.; Tagliaferri, G.; Tenzer, C.; Vink, J.; Weisskopf, M.C.; Zane, S.; Agudo, I.; Antonelli, A.; Attina, P.; Baldini, L.; Bykov, A.; Carpentiero, R.; Cavazzuti, E.; Churazov, E.; Del Monte, E.; De Martino, D.; Donnarumma, I.; Doroshenko, V.; Evangelista, Y.; Ferreira, I.; Gallo, E.; Grosso, N.; Kaaret, P.; Kuulkers, E.; Laranaga, J.; Latronico, L.; Lumb, D.H.; Macian, J.; Malzac, J.; Marin, F.; Massaro, E.; Minuti, M.; Mundell, C.; Ness, J.U.; Oosterbroek, T.; Paltani, S.; Pareschi, G.; Perna, R.; Petrucci, P.O.; Pinazo, H.B.; Pinchera, M.; Rodriguez, J.P.; Roncadelli, M.; Santovincenzo, A.; Sazonov, S.; Sgro, C.; Spiga, D.; Svoboda, J.; Theobald, C.; Theodorou, T.; Turolla, R.; Wilhelmi de Ona, E.; Winter, B.; Akbar, A.M.; Allan, H.; Aloisio, R.; Altamirano, D.; Amati, L.; Amato, E.; Angelakis, E.; Arezu, J.; Atteia, J.L.; Axelsson, M.; Bachetti, M.; Ballo, L.; Balman, S.; Bandiera, R.;

Barcons, X.; Basso, S.; Baykal, A.; Becker, W.; Behar, E.; Beheshtipour, B.; Belmont, R.; Berger, E.; Bernardini, F.; Bianchi, S.; Bisnovatyi-Kogan, G.; Blasi, P.; Blay, P.; Bodaghee, A.; Boer, M.; Boettcher, M.; Bogdanov, S.; Bombaci, I.; Bonino, R.; Braga, J.; Brandt, W.; Brez, A.; Bucciantini, N.; Burderi, L.; Caiazzo, I.; Campana, R. XIPE: the x-ray imaging polarimetry explorer. Space Telescopes and Instrumentation 2016: Ultraviolet to Gamma Ray, 2016, Vol. 9905, Proceedings of SPIE, p. 990515.

40. Zhang, S.N.; Feroci, M.; Santangelo, A.; Dong, Y.W.; Feng, H.; Lu, F.J.; Nandra, K.; Wang, Z.S.; Zhang, S.; Bozzo, E.; Brandt, S.; De Rosa, A.; Gou, L.J.; Hernanz, M.; van der Klis, M.; Li, X.D.; Liu, Y.; Orleanski, P.; Pareschi, G.; Pohl, M.; Poutanen, J.; Qu, J.L.; Schanne, S.; Stella, L.; Uttley, P.; Watts, A.; Xu, R.X.; Yu, W.F.; in 't Zand, J.J.M.; Zane, S.; Alvarez, L.; Amati, L.; Baldini, L.; Bambi, C.; Basso, S.; Bhattacharyya S..; , R., B.; Belloni, T.; Bellutti, P.; Bianchi, S.; Brez, A.; Bursa, M.; Burwitz, V.; Budtz-Jørgensen, C.; Caiazzo, I.; Campana, R.; Cao, X.; Casella, P.; Chen, C.Y.; Chen, L.; Chen, T.; Chen, Y.; Chen, Y.; Chen, Y.P.; Civitani, M.; Coti Zelati, F.; Cui, W.; Cui, W.W.; Dai, Z.G.; Del Monte, E.; de Martino, D.; Di Cosimo, S.; Diebold, S.; Dovciak, M.; Donnarumma, I.; Doroshenko, V.; Esposito, P.; Evangelista, Y.; Favre, Y.; Friedrich, P.; Fuschino, F.; Galvez, J.L.; Gao, Z.L.; Ge, M.Y.; Gevin, O.; Goetz, D.; Han, D.W.; Heyl, J.; Horak, J.; Hu, W.; Huang, F.; Huang, Q.S.; Hudec, R.; Huppenkothen, D.; Israel, G.L.; Ingram, A.; Karas, V.; Karelin, D.; Jenke, P.A.; Ji, L.; Korpela, S.; Kunneriath, D.; Labanti, C.; Li, G.; Li, X.; Li, Z.S.; Liang, E.W.; Limousin, O.; Lin, L.; Ling, Z.X.; Liu, H.B.; Liu, H.W.; Liu, Z.; Lu, B.; Lund, N.; Lai, D.; Luo, B.; Luo, T.; Ma, B.; Mahmoodifar, S.; Marisaldi, M.; Martindale, A.; Meidinger, N.; Men, Y.; Michalska, M.; Mignani, R.; Minuti, M.; Motta, S.; Muleri, F.; Neilsen, J.; Orlandini, M.; Pan, A.T.; Patruno, A.; Perinati, E.; Picciotto, A.; Piemonte, C.; Pinchera, M.; Rachevski A..; Rapisarda, M.; Rea, N.; Rossi, E.M.R.; Rubini, A.; Sala, G.; Shu, X.W.; Sgro, C.; Shen, Z.X.; Soffitta, P.; Song, L.; Spandre, G.; Stratta, G.; Strohmayer, T.E.; Sun, L.; Svoboda, J.; Tagliaferri, G.; Tenzer, G.; Hong, T.; Taverna, R.; Torok, G.; Turolla, R.; Vacchi, S.; Wang, J.; Walton, D.; Wang, K.; Wang, J.F.; Wang, R.J.; Wang, Y.F.; Weng, S.S.; Wilms, J.; Winter, B.; Wu, X.; Wu, X.F.; Xiong, S.L.; Xu, Y.P.; Xue, Y.Q.; Yan, Z.; Yang, S.; Yang, X.; Yang, Y.J.; Yuan, F.; Yuan, W.M.; Yuan, Y.F.; Zampa, G.; Zampa, N.; Zdziarski, A.; Zhang, C.; Zhang, C.L.; Zhang, L.; Zhang, X.; Zhang, Z.; Zhang, W.D.; Zheng, S.J.; Zhou, P.; Zhou X. L.. eXTP: Enhanced X-ray Timing and Polarization mission. Space Telescopes and Instrumentation 2016: Ultraviolet to Gamma Ray, 2016, Vol. 9905, Proceedings of SPIE, p. 99051Q, [arXiv:astro-ph.IM/1607.08823].

41. De Angelis, A.; Tatischeff, V.; Tavani, M.; Oberlack, U.; Grenier, I.; Hanlon, L.; Walter, R.; Argan, A.; von Ballmoos, P.; Bulgarelli, A.; Donnarumma, I.; Hernanz, M.; Kuvvetli, I.; Pearce, M.; Zdziarski, A.; Aboudan, A.; Ajello, M.; Ambrosi, G.; Bernard, D.; Bernardini, E.; Bonvicini, V.; Brogna, A.; Branchesi, M.; Budtz-Jorgensen, C.; Bykov, A.; Campana, R.; Cardillo, M.; Coppi, P.; De Martino, D.; Diehl, R.; Doro, M.; Fioretti, V.; Funk, S.; Ghisellini, G.; Grove, E.; Hamadache, C.; Hartmann, D.H.; Hayashida, M.; Isern, J.; Kanbach, G.; Kiener, J.; Knödlseder, J.; Labanti, C.; Laurent, P.; Limousin, O.; Longo, F.; Mannheim, K.; Marisaldi, M.; Martinez, M.; Mazziotta, M.N.; McEnery, J.; Mereghetti, S.; Minervini, G.; Moiseev, A.; Morselli, A.; Nakazawa, K.; Orleanski, P.; Paredes, J.M.; Patricelli, B.; Peyré, J.; Piano, G.; Pohl, M.; Ramarijaona, H.; Rando, R.; Reichardt, I.; Roncadelli, M.; Silva, R.; Tavecchio, F.; Thompson, D.J.; Turolla, R.; Ulyanov, A.; Vacchi, A.; Wu, X.; Zoglauer, A. The e-ASTROGAM mission. Exploring the extreme Universe with gamma rays in the MeV - GeV range. Experimental Astronomy **2017**, 44, 25–82, [arXiv:astro-ph.HE/1611.02232].

42. Tatischeff, V.; De Angelis, A.; Gouiffès, C.; Hanlon, L.; Laurent, P.; Madejski, G.; Tavani, M.; Ulyanov, A. e-ASTROGAM mission: a major step forward for gamma-ray polarimetry. ArXiv e-prints **2017**, [arXiv:astro-ph.HE/1706.07031].

43. De Angelis, A.; Tatischeff, V.; Grenier, I.A.; McEnery, J.; Mallamaci, M.; Tavani, M.; Oberlack, U.; Hanlon, L.; Walter, R.; Argan, A.; et al.. Science with e-ASTROGAM (A space mission for MeV-GeV gamma-ray astrophysics). ArXiv e-prints **2017**, [arXiv:astro-ph.HE/1711.01265].